\documentclass{IEEEcsmag}

\usepackage[colorlinks,urlcolor=blue,linkcolor=blue,citecolor=blue]{hyperref}
\expandafter\def\expandafter\UrlBreaks\expandafter{\UrlBreaks\do\/\do\*\do\-\do\~\do\'\do\"\do\-}
\usepackage{upmath,color}
\usepackage[super,comma]{natbib}

\usepackage{multicol}

\usepackage{etoolbox}

\jvol{XX}
\jnum{XX}
\paper{8}
\jmonth{Month}
\jname{IEEE Security and Privacy}
\jtitle{Publication Title}
\pubyear{2024}

\begin{document}

\sptitle{Feature Article - Achieving Autonomous Cyber Defense}

\title{The Path To Autonomous Cyber Defense}

\author{Sean Oesch}
\affil{Oak Ridge National Laboratory, Oak Ridge, TN, USA, oeschts@ornl.gov}

\author{Phillipe Austria}
\affil{Oak Ridge National Laboratory, austriaps@ornl.gov}

\author{Amul Chaulagain}
\affil{Oak Ridge National Laboratory, chaulagaina@ornl.gov}

\author{Brian Weber}
\affil{Oak Ridge National Laboratory, weberb@ornl.gov}

\author{Cory Watson}
\affil{Oak Ridge National Laboratory, watsoncl1@ornl.gov}

\author{Matthew Dixson}
\affil{Oak Ridge National Laboratory, dixsonmk@ornl.gov}

\author{Amir Sadovnik}
\affil{Oak Ridge National Laboratory, sadovnika@ornl.gov}

\markboth{THEME/FEATURE/DEPARTMENT}{THEME/FEATURE/DEPARTMENT}

\begin{abstract}
\looseness0-Defenders are overwhelmed by the number and scale of attacks against their networks.This problem will only be exacerbated as attackers leverage artificial intelligence to automate their workflows. We propose a path to autonomous cyber agents able to augment defenders by automating critical steps in the cyber defense life cycle. 
\end{abstract}

\maketitle

\chapteri{T}he proliferation of professionalized cyber crime groups combined with the shortage of skilled professionals in cyber defense has left defenders incapable of handling the volume of new attacks targeting their networks.\footnote{World Economic Forum, Global Cybersecurity Outlook 2023. \url{https://www.weforum.org/reports/global-cybersecurity-outlook-2023/}, July 13, 2023.}  
To avoid being overwhelmed, defenders must automate security policies, detection, and mitigation, yet many of these tasks still require a large time investment from human personnel. 
The creation of autonomous cyber defense agents is one promising approach to automate operations and prevent cyber defenders from being overwhelmed. 
This article lays out the steps that researchers and practitioners need to take on the path to practical autonomous cyber defense agents, with a focus on reinforcement learning as a promising approach.\footnote{This work has been submitted to the IEEE for possible publication. Copyright may be transferred without notice, after which this version may no longer be accessible.}

Reinforcement learning (RL) addresses the challenge of “learning from interaction with an environment in order to achieve long-term goals”, where “long-term goals” could include flying a helicopter safely to the intended destination, winning a game of chess, efficiently managing a power station, or protecting a network against cyber attacks. At a high level, reinforcement learning agents map situations (a state) to actions via a “policy” function in order to maximize a numerical reward signal. Rewards may vary in frequency (a single win/loss reward at the end of a game or a repeated “distance from goal” reward while a robot is walking) and complexity.

In order to handle large state spaces, deep reinforcement learning (deep RL) takes advantage of deep neural nets in order to generalize well across states. By leveraging deep RL, DeepMind  has trained reinforcement learning algorithms to defeat expert human players in games such as Go\footnote{\url{https://www.deepmind.com/research/highlighted-research/alphago}}  and StarCraft\footnote{\url{https://www.deepmind.com/blog/alphastar-mastering-the-real-time-strategy-game-starcraft-ii}}. When cybersecurity is modeled as a game between a defender (blue team) and an attacker (red team) it has similarities to games like StarCraft, where the state space and action space are large (\begin{math}10^{26}\end{math} for StarCraft). Because reinforcement learning has demonstrated the ability to defeat human adversaries in complex games with large state spaces, it is a natural choice for creating defensive cyber agents.

As a result of this parity between cybersecurity and other areas where RL has proven effective, both cyber defenders and attackers have sought to leverage this potential. 
Could autonomous RL agents be used to help defenders delay and deny attackers?~\cite{hammar2020finding,janisch2023nasimemu,ridley2018machine,nyberg2023training,walter2021incorporating,dutta2023deep}
Could autonomous RL agents be leveraged by defenders to automate pen testing?~\cite{hu2020automated}
Could autonomous RL agents be leveraged by attackers to overwhelm or sneak past defenders?
In order to help answer these questions, researchers developed simulation~\cite{team2021cyberbattlesim,nyberg2023training,hammar2020finding,dutta2023deep} and emulation environments in which to train such autonomous RL agents, with some environments~\cite{janisch2023nasimemu,ridley2018machine,molina2021network} providing both simulation and emulation. 

Yet in spite of prior efforts to create autonomous cyber agents, they are still not used in practice. 
What is preventing the adoption of this new and promising approach to cyber defense? 
This article seeks to answer this question by highlighting several of the challenges faced by researchers in this field and laying out the path to autonomous cyber defense. 
We begin by situating the possible roles that autonomous agents can play within the context of the cyber defense life cycle and then discuss the need for a properly defined "game" for the agent to play, the need for agent adaptability to changing network environments and evolving adversary behaviors, and the need for better training environments to facilitate future research. 
\vspace*{-5pt} 

\section{THE CYBER DEFENSE LIFE CYCLE}
The National Institute of Standards and Technology (NIST) has defined 6 key stages in their cybersecurity framework\footnote{\url{https://nvlpubs.nist.gov/nistpubs/CSWP/NIST.CSWP.29.pdf}}: 

\begin{itemize}
    \item \textit{Govern}: The policies and risk management strategy of the organization.
    \item \textit{Identify}: Knowing the organizations current cybersecurity risks. 
    \item \textit{Protect}: Safeguards that help manage existing risks (identity management, authentication, access control). 
    \item \textit{Detect}: Find and analyze potential threats to the network. 
    \item \textit{Respond}: Responding to a threat once it has been identified.
    \item \textit{Recover}: Restoring assets impacted by the threat after it has been neutralized. 
\end{itemize}

In the context of these 6 stages of the cyber defense life cycle, what role or roles should autonomous agents play? 
Current training environments for autonomous cyber defense agents give the agent a set of actions that span detection (the agent is fed host and network logs and is able to create honeypots or honey networks), response (the agent can add firewall rules, block IPs, or isolate hosts), and recovery (the agent can restore boxes to their original state). 
Yet little work has been done to understand if these actions would work in a real cyber defense context or if these are even the actions that we should be automating. 
More research is needed to understand the types of autonomous agents that would most benefit cyber analysts in their existing workflows, which automated actions would be permitted based on the policy documents governing Security Operations Centers (SOCs), how to avoid replicating existing commercial tools, and the practicality of training these agents at scale over different action and observation spaces. 

Within the context of the cyber defense life cycle, it is also unclear that a single autonomous agent should be responsible for multiple stages of the life cycle. 
Current methods for creating autonomous agents tend to perform better when the number of actions available to them is smaller.
In cybersecurity, one practical way to reduce the number of actions that an agent is required to perform is to limit the agent to only a single stage of the cyber defense life cycle, and in some cases only a specific function within that stage. 
Rather than creating a single monolithic agent, it would be better to identify which portions of each stage of the life cycle should be automated and then create multiple autonomous agents, each with a specialization, that work together to achieve the larger goal of defending the network autonomously. 
These agents can then work in tandem or independently.

We believe that such a multi-agent approach is likely the best path forward to create reliable autonomous agents for cyber defense because it makes both creation and adoption easier. 
An autonomous agent with a more specific goal and smaller action space is easier to create, test, and deploy. 
Such agents make the overall system more modular, so that individual agents can be swapped out or removed with minimal impact to the overall performance of the system. 
It is also easier for a SOC to adopt because it will require less effort to integrate into their existing toolset and it will require less input data which a given SOC may or may not currently be collecting. 

\section{PLAYING THE RIGHT GAME}
One of the ingredients necessary for the successful creation of autonomous cyber defense agents is that the agent must be playing the right game. 
In many contexts where autonomous agents succeed there is a well defined game with clear boundaries, goals, and end states. 
For example, when piloting a drone the goal is to get from point A to point B without hitting any obstacles, perhaps with a few sub-goals such as staying within certain elevation boundaries, avoiding no go zones, and/or delivering a package. 
In literal games like chess, Go, or StarCraft, the rules are even more well defined and the environments are somewhat predictable. 

Cybersecurity cannot be reduced to a single game and the environment in which the agent operates may change dynamically (not to mention that adversaries constantly seek to rewrite the rules of the game). 
Defining the correct game for autonomous cyber agents therefore presents two critical challenges: there is no single game and the game itself changes. 
An additional challenge is that different organizations' networks may vary in terms of scale, topology, support staff, services, and defensive tools deployed. 
The consequence of this variability is that any autonomous agent must be able to adapt to a new environment even as it must be able to adapt to shifts in the game itself. 
The \textit{right game} is therefore difficult to define in a cybersecurity context. 

To further illustrate this point, let us consider a specific example of how even small details in agent design impact agent usability and performance. 
The Cyber Operations Research Gym (CybORG)~\cite{standen2021cyborg} is a training environment for autonomous agents.
We will focus on the version of CybORG provided in the second Cyber Autonomy Gym for Experimentation (CAGE) challenge~\footnote{\url{https://github.com/cage-challenge/cage-challenge-2}}, which focused specifically on defending an enterprise network.
More specifically, we will examine the design of the observation space (the set of information the autonomous agent uses to make decisions) in CybORG.
The structure of the observation space is critical to agent usability and performance. 

The CybORG simulator maintains a detailed global network state, containing information on each individual host such as its IP address, active connections, running services, etc. 
The Blue and Red agent take turns running actions on the network based on this state. 
The default observation space is a filtered view of the full simulated state of the network. 
It is a 1D vector where each host is assigned 2 fields: Activity State, and Compromised State.
The Activity State indicates whether a red action was taken on a given host in the previous turn, and can be three states: None (no red action), Scan (a red agent discovered the host), and Exploit (a red agent exploited the host).
The Compromised State shows whether a host has been compromised and to what extent it is compromised.

\begin{center}
\begin{tabular}{|c | c | c|} 
    \hline
    Hosts & Activity & Compromised \\
    \hline
    Host1 & None & No \\
    \hline
    Host2 & Exploit & Privileged \\
    \hline
    ... & ... & ... \\
    \hline
\end{tabular}
\end{center}

\begin{figure*}
    \centering
    \hbox{\includegraphics[scale = .38]{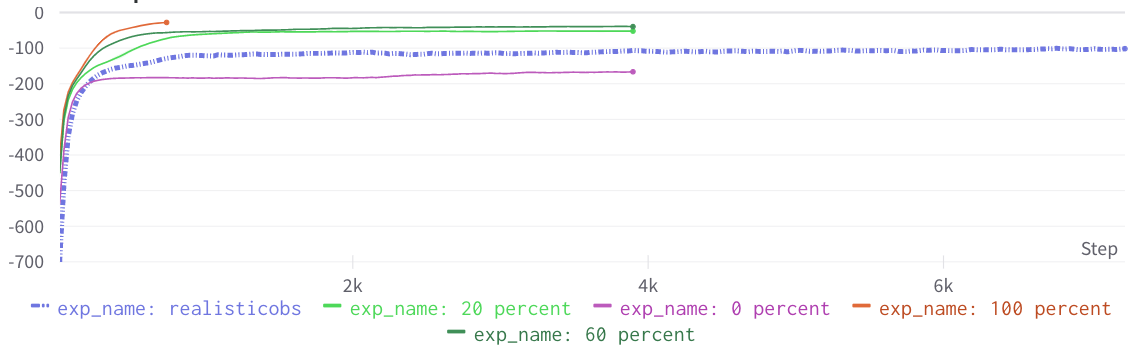}}
    \caption{When training an autonomous agent, the probabilities of observing different attacker behaviors need to match real life probabilities or else the agent may fail in a real environment.
    The dotted blue line utilized what we considered realistic detection probabilities for specific red agent actions (50\% for adding a new user, 15\% for adding a process, 5\% adding new session, etc.)
    Defining observation probabilities in only a small part of creating the appropriate game environment. 
    }
    \label{fig:stochvreal}
\end{figure*}

By default the blue agent always observes the red agent actions in CybORG. 
In other words, the blue agent has perfect visibility into what the red agent is doing. 
As part of our research on CybORG we tested the simulator with different parameters such as network size, attacker strategies, and the training algorithm. 
We also extended CybORG to enable new capabilities to support our research. 
For example, we added the ability to configure the probability of red agent actions being detected by the blue agent. 
Our results, shown in Figure~\ref{fig:stochvreal}, highlight the impact of detection probabilities on agent performance. 
The probabilities of observing different attacker behaviors need to match real life probabilities or else the agent may fail in a real environment. 
Defining detection probabilities is only a small piece of defining the correct game to play when training autonomous cyber defense agents. 

Yet there is an issue more fundamental than detection probabilities with the observation space defined by CybORG for the second CAGE challenge. 
The agent is ingesting raw log data. 
Existing cybersecurity tools already ingest this data and produce alerts for intrusions at the host and network level. 
The agent trained in CybORG is therefore replicating existing capabilities and trying to learn new behaviors on top of those existing capabilities. 
How might we change the observation space to avoid replicating existing tools' capabilities? 

The MITRE ATT\&CK framework is used to define both red agent actions and the data sources collected by sensors / logs relevant to detecting each type of attack. 
These data sources are called Data Components\footnote{https://attack.mitre.org/datasources/} and they are used to determine probability of detecting specific red actions. 
For each Data Component that tactic or technique touches, we could define a detector that is then given a probability of detecting a specific red agent action. 

\begin{figure*}
    \centering
    \hbox{\hspace{-1em}\includegraphics[scale = .37]{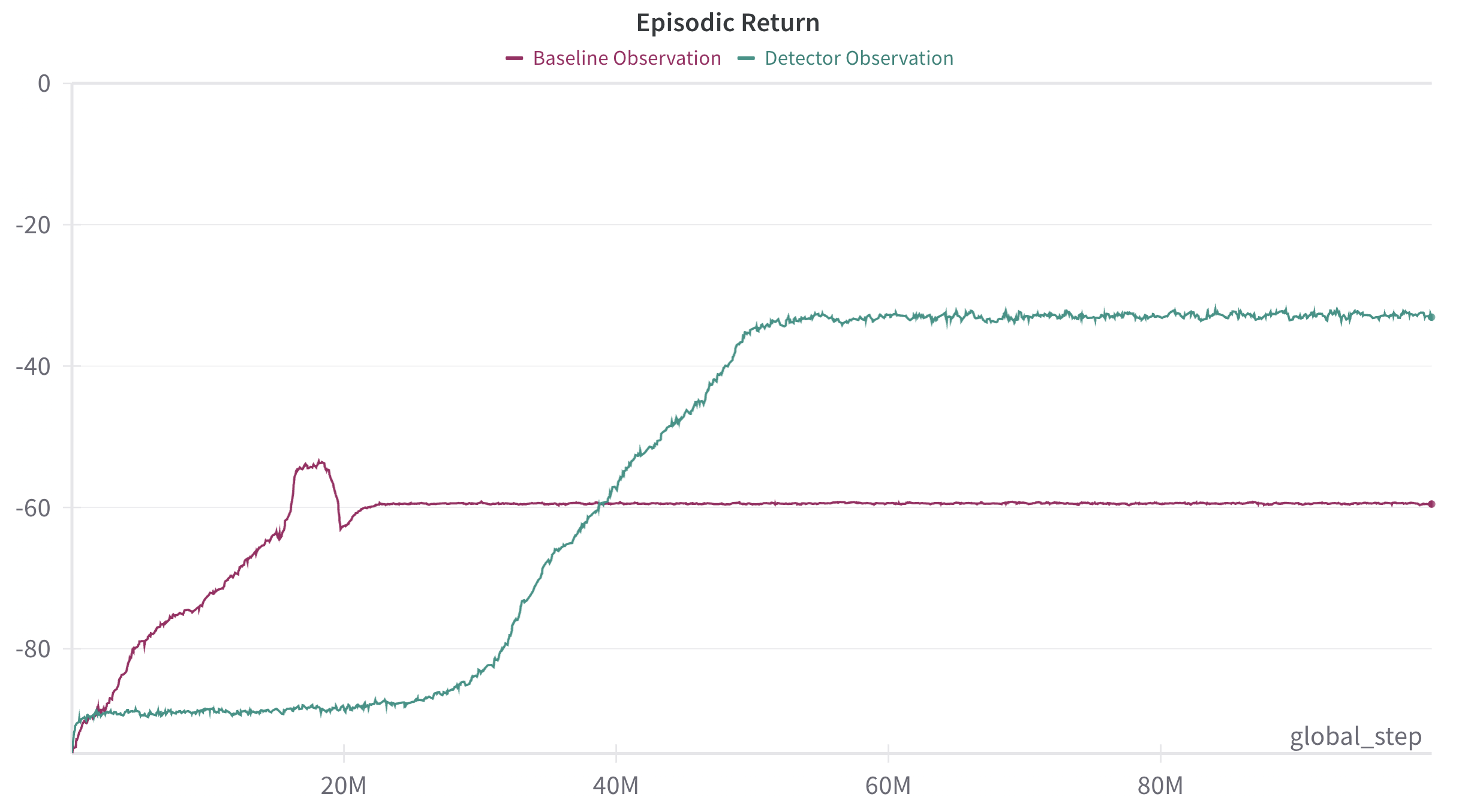}}
    \caption{The new detector based observation implementation with a perfect detector outperforms the baseline (original CybORG observation space) over 100M training steps with 10 subnets, demonstrating the feasibility of utilizing an alternate observation space that avoids replicating existing tools and is easier to use in existing SOCs.
    This new observation space represents a more usable action space both in terms of integration for SOCs and from the perspective of the reinforcement learning algorithm.
    When training autonomous cyber agents, every component of the game, from the observation space to the reward and actions, must be appropriately defined for the agent to be usable after training.}
    \label{fig:obs_comparison}
\end{figure*}

When a red action occurs, the detector ``flips a coin'' based on the probabilities of detection for each Data Component. 
If any flip signifies success, the alert is fed as an observation to the blue agent. 
Each detector would also have an object for describing false positive generation probability. 
By utilizing detectors as the signal rather than network or host logs, we provide the agent with a slightly noisier yet more realistic signal, avoid replicating the capabilities of existing (likely more accurate) detection tools, all while improving flexibility and simplifying deployment to any network containing those existing cyber-detection tools.
We implemented this new observation inside CybORG and Figure~\ref{fig:obs_comparison} shows that this new observation space is just as effective as the original observation space. 

We are not suggesting that this alternate observation space is necessarily the optimal observation space. 
Rather, the takeaway is that it is critical to define realistic state spaces that will transfer to operational networks more effectively in practice.
And even more critically that we must define the \textit{right game}, from the observation space to the reward function and actions, in order for autonomous cyber defense agents to be usable and effective in a real network. 

\section{ADAPTABILITY IS KEY}
In addition to playing the right game, an autonomous cyber defense agent needs to be adaptable in the following ways: 

\begin{enumerate}
    \item Be adaptable to varying networks. 
    Networks vary greatly across organizations and constantly change as devices join/leave the network.
    \item Be adaptable to varying adversary strategies. 
    Adversaries have varying goals that change over time and utilize different tactics, techniques, and procedures (TTPs).
    \item Be adaptable to different objectives in the CIA (confidentiality, integrity, availability) triangle. 
    Companies may prioritize different objectives depending on what most impacts their bottom line and the objectives they value most may fluctuate with time. 
\end{enumerate}

One of the reasons that adaptability is critical is that the resources necessary to train one of these agents are significant. 
The current algorithms used to train these agents require thousands if not hundreds of thousands or millions of rounds of training in which the environment begins in a known starting state, the game is played to completion, and then the environment is reset to the starting state (with some stochasticity built in for variability where desired). 
Such training is not possible in a real network given the amount of time it would take to reconfigure the environment (not to mention initial setup). 
It is therefore necessary to use either a simulated or emulated environment, with a simulated environment being the most practical because it allows rapid retraining of agents as networks, adversaries, and organizations shift over time. 
Ideally an agent would not need such retraining and could adapt on its own, but an intermediate step towards that goal may be rapid retraining of agents as necessary. 

The sheer cost of setting up simulation/emulation environments can be seen in the approach of DARPA's Cyber Agents for Security Testing and Learning Environments (CASTLE) program\footnote{\url{https://www.darpa.mil/program/cyber-agents-for-security-testing-and-learning-environments}}, which utilizes separate contractors to design the simulation and emulation environments, as well as another to design the actual agents. 
In addition to the simulation and emulation environment, CASTLE also requires the companies designing the agents to show that they work inside of an actual network, which they plan to achieve by utilizing a subsection of an existing network. 
Yet even after demonstrating that an autonomous agent trained in simulation and emulation works in a real network, that does not mean that it will be trivial to retrain the agent to work in a different organization's network where the available observation data, topology, or goals are different. 
Because training these agents is resource intensive, it is critical that these agents be designed from the ground up to be adaptable to different environments with minimal retraining. 

Another area where agents need to be adaptable is in regard to changing adversary behaviors. 
The ability to generalize to new strategies is nontrivial for existing autonomous agent algorithms such as reinforcement learning (RL). 
RL agents can learn to do complex tasks in a specific environment, but often fail to complete the same tasks even in environments that are largely similar. 
Many RL agents are tested on the same environment they were trained on, which in traditional machine learning would be the same as testing on your training data.\footnote{\url{https://openai.com/research/quantifying-generalization-in-reinforcement-learning}}

As a result of the limitations of existing algorithms for autonomous agent design, novel approaches are needed to address the challenge of adaptability, both to new adversary behaviors as well as new network contexts. 
It is difficult to see how this form of adaptability can be achieved with current algorithms without continuous retraining of the agent, either in the actual environment or in a shadow environment. 
Current deep neural network based cyber solutions do utilize online training to adapt to new data and avoid problems such as concept drift, but in the case of autonomous agents the cost of continuous retraining would be higher than with existing technologies. 
This problem of adaptability is an area ripe for future research into ways to either make the agents more generalizable or retrain more rapidly as conditions change. 

In order to highlight this issue it is useful to describe how RL algorithms work, why adaptability is a challenge, and how more recent works attempt to deal with this challenge. Although RL algorithms have been around for decades, they had a resurgence (along with the rest of the artificial intelligence field) with the rise of deep learning. Combining the two formed a new set of methods called deep reinforcement learning (DRL), which combine the power of deep learning methods to learn complex non-linear functions and generalize over inputs by learning important features and the ability of RL to learn from interactions with an environment without the need for ground truth data. This led to many state-of-the-art results in games such as Atari \cite{mnih2013playing} an Go \cite{silver2016mastering}.

This makes DRL an ideal candidate to use as an autonomous cyber defense agent. However, in spite of the excellent results DRL can achieve it also has certain characteristics which limit its capability to adapt. First, deep neural networks usually rely on constant size inputs and outputs. In our case this means that the size of the observation space needs to be constant (which is a challenge if we want the agent to be able to deal with different size networks) as well as the number of actions needs to remain constant (a challenge if we want to add a new defensive action given a new capability). Changing either of those would usually require a complete retraining of the DRL network. The second issue is DL's reliance on massive amounts of data, which in an RL setting means many experiences.Therefore any change to the environment or agent capabilities could end up being computationally prohibitive.

Given these issues developing more adaptive RL is an active area of research which should be adopted by the autonomous cyber agents research community. For example, \cite{pmlr-v119-jain20b} proposes a two stage framework to be able to generalize to new actions. That is, the first stage infers a representation for the action (including a new action) which then allows it to be used in the second stage. Although, they show the usefulness of this method in some simple games, it is unclear how well this method will work in the cyber domain. Variable observation space can be handled in multiple ways. For example, input can be ``summarized'' using a transformer architecture which can deal with variable input sizes. Although this has proven useful for domains such as text \cite{9231622} it is still needs to be better investigated for the cyber domain.

One area where existing algorithms may be sufficient to provide adaptability is in supporting different organization goals. 
Autonomous agents based on RL use a reward function to tell the agent what outcomes are important. 
By using a dynamic reward functions that allow defenders to tweak configuration parameters it should be possible to prioritize different points in the CIA triangle without retraining by changing the weight assigned to each goal.
One organization may assign a greater weight to availability, while another may assign a greater weight to confidentiality. 
For example, consider the following priorities and examples of organizations who would have those priorities. 

\begin{enumerate}
    \item Prioritize confidentiality. Ex: PCI (Payment Card Industry) infrastructure.
    \item Prioritize trapping red actors in honey networks to study behavior. Ex: FireEye/Research/IT Company with mature environment.
    \item Prioritize availability in favor of removing attackers or moving to honey network as long as service is maintained. Ex: Wikipedia / web sharing service.
\end{enumerate}

\section{Better Training Environments}
Putting it all together, we now discuss what is required to create better training environments to train autonomous agents for cyber defense. 
As noted by Vyas et al.~\cite{vyas2023automated}, prior efforts towards automated cyber defense training environments fall short of providing the level of fidelity necessary to train agents that can generalize well or be efficiently transferred to an operational network. 
Given the importance of adaptability highlighted in the previous section, we need to create training environments capable (given the right algorithms) of producing autonomous agents that are (a) Adaptable to a variety of network sizes and topologies and (b) Adaptable to varying adversary strategies. 
At a minimum, such a training environment should provide the following: 

\begin{enumerate}
    \item An observation space based on signals available in operational networks with tunable detection probabilities and the ability to simulate / emulate the impact of false positives and false negatives on performance.
    \item The ability to dynamically generate realistic varying network topologies and individual host configurations in the network.
    \item The ability to support the training of autonomous agents that utilize different algorithms and libraries with minimal reconfiguration. 
    \item A visualization tool that enables autonomous agent creators to diagnose and understand agent behavior by replaying and observing episodes.
    \item An emulation environment that is extensible, scalable, repeatable, and allows the evaluation of agent transferability to actual networks.
\end{enumerate}

In addition to the shortcomings of available environments, there is also the problem of inconsistency. 
Because each environment represents the network, reward, and observation space differently, it is not possible to compare the results obtained across environments. 
The problem of inconsistency is separate from the issue of realism, where the environment does not match real network behavior closely enough to train a transferable agent. 
A standardized set of environments for training that allow researchers to focus on developing the science of autonomous agents for cyber rather than having to create their own experimental environment from scratch, which may take a year or more even when the resulting environment is not realistic enough to produce agents that can be transferred to a  real network, is critical to the progress of the field of autonomous cyber defense. 

\begin{figure}
    \centering
    \hbox{\includegraphics[scale = .22]{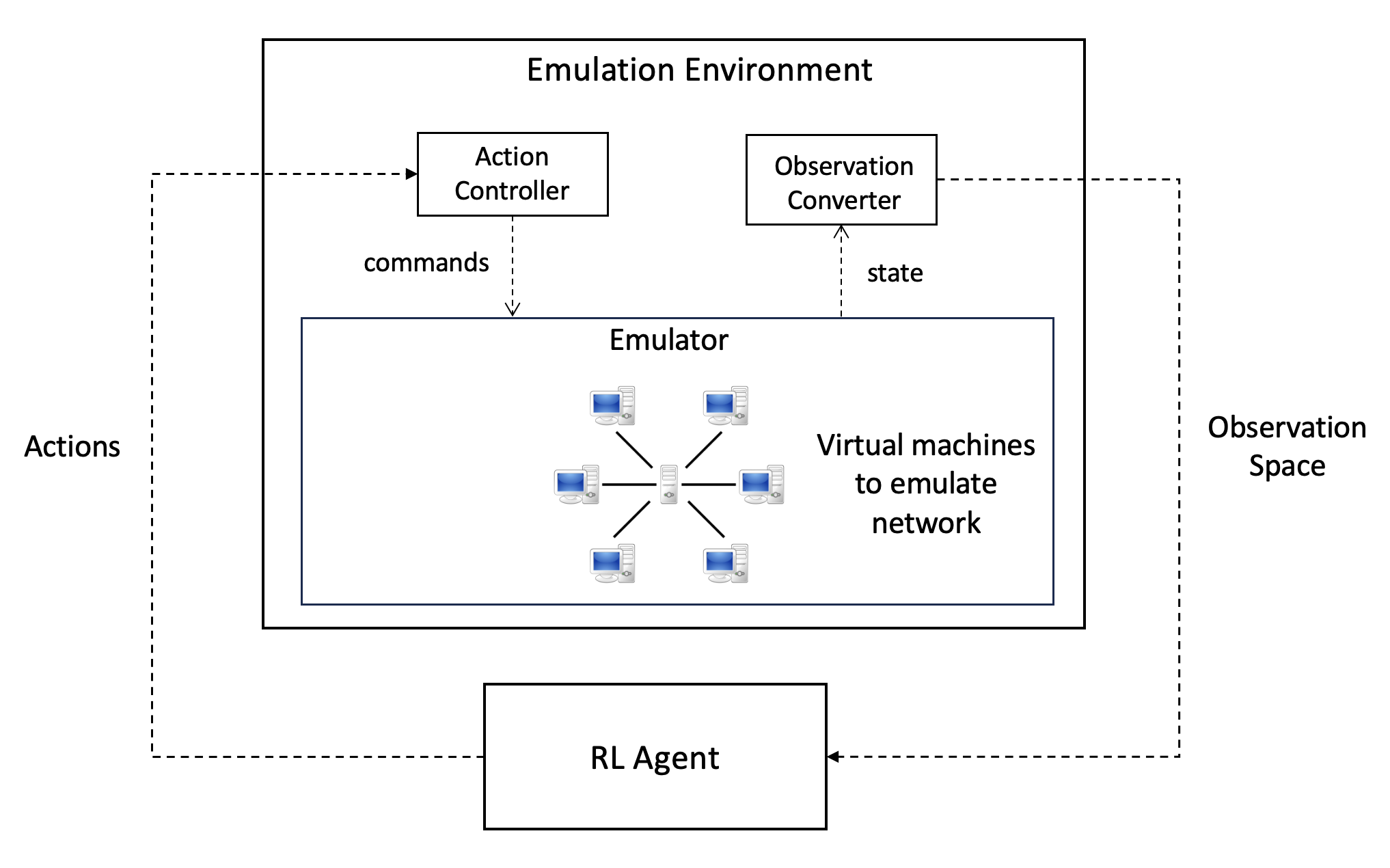}}
    \caption{The emulation environment consists of three main components: the Action Controller, Observation Converter and Emulator. 
    The Action Controller receives actions from the RL Agent and transmits commands to the Emulator, while the Observation Converter converts the state of the emulator to an observation space vector to be fed back to the RL Agent. 
    The Emulator manages virtual machines that represent hosts and other entities in a network.
    Any changes to the set of available actions for either offense or defense, the observation space, or the emulation environment require significant time investment, and maintaining all of the tools necessary to run each of the components is also labor intensive.}
    \label{fig:emulation}
\end{figure}

Part of what makes creating these research environments so challenging is the need for both simulation and emulation.
Writing a high fidelity simulated environment is difficult. 
Creating an environment for emulation that contains all of the necessary actions, observations, and behaviors is harder still because every action must be integrated into the network rather than merely represented by code. 
Figure \ref{fig:emulation} illustrates how an emulation environment interacts with a trained RL Agent and highlights the complexity of creating, updating, and maintaining these environments. 

The amount of work intensive efforts required to develop and deploy an emulation environment has limited the availability of related research and open source options.
Some challenges with emulation we foresee to overcome are:

\begin{itemize}
    \item Being able to evaluate the agent in multiple scenarios that vary network size and complexity, especially without having to perform considerable amounts of manual reconfiguration.
    \item Developing a system to collect the necessary state information required by the agent and properly converting to an observation space.
    \item Reducing latency between the emulator and agent to ensure smooth evaluation execution.
\end{itemize}

\vspace*{-5pt} 
 
\section{CONCLUSION}
We believe that the path to autonomous cyber defense will be multi-agent solutions where agents specialize in specific areas of the cyber defense life cycle. 
Training these agents will require high fidelity training environments that are openly available so that researchers can directly compare their results and new algorithms capable of adapting to shifts in network topology, adversary behavior, and organizational priorities.
Because training these agents is resource intensive, it is critical that these agents be designed from the ground up to be adaptable to different environments with minimal retraining.
Early steps toward this goal will likely involve training environments that allow organizations to rapidly retrain deployed agents to adapt to new threats. 

Much work still needs to be done to understand the best applications for these agents and how to define the observation space, reward function, and environments for training in each application space. 
For further reading or to experiment with different simulation and emulation environments, we suggest the following GitHub page, which contains links to papers and existing environments: ~\url{https://github.com/Limmen/awesome-rl-for-cybersecurity}

\vspace*{-8pt}

\section{ACKNOWLEDGMENTS}
Notice: This manuscript has been authored by UT-Battelle, LLC under Contract No. DE-AC05-00OR22725 with the U.S. Department of Energy. The United States Government retains and the publisher, by accepting the article for publication, acknowledges that the United States Government retains a non-exclusive, paid-up, irrevocable, world-wide license to publish or reproduce the published form of this manuscript, or allow others to do so, for United States Government purposes. The Department of Energy will provide public access to these results of federally sponsored research in accordance with the DOE Public Access Plan (http://energy.gov/downloads/doe-public-access-plan).

This manuscript was prepared as part of the Emerging and Cyber Security Technologies initiative at Oak Ridge National Laboratory. 

\def\refname{REFERENCES}

\bibliographystyle{plainnat}
\bibliography{references}

\begin{IEEEbiography}{Sean Oesch}{\,}is a researcher at Oak Ridge National Laboratory, Oak Ridge, Tennessee. His current research interests include autonomous cyber defense, explainable AI for cyber applications, and AI for cyber defense. Sean received his Ph.D. from the University of Tennessee, Knoxvile, where he studied password manager security. He is a Senior Member of IEEE. \vspace*{8pt}
\end{IEEEbiography}

\begin{IEEEbiography}{Phillipe Austria}{\,} is a researcher at Oak Ridge National Laboratory, Oak Ridge, Tennessee. His currently interest include data visualizations, machine learning and blockchain . Phillipe received his Ph.D. from the University of Nevada, Las Vegas, where he studied computer science with a focus on decentralized systems.\vspace*{8pt}
\end{IEEEbiography}

\begin{IEEEbiography}{Amul Chaulagain}{\,}is a software engineer working at Oak Ridge National Laboratory. He is currently involved in projects involving the utilization of autonomous systems and AI for cyber defense. Amul received his B.S. from Purdue University, where he studied Computer Science with a concentration on Software Engineering.
\end{IEEEbiography}

\begin{IEEEbiography}{Matthew Dixson}{\,} works at Oak Ridge National Laboratory, Oak Ridge, Tennessee as a cybersecurity technical professional. He is currently interested in applying AI to cuber defense. Matthew received his M.S. from the University of Tennessee, Knoxville where he studied computer science with a concentration in cybersecurity.  
\end{IEEEbiography}

\begin{IEEEbiography}{Amir Sadovnik}{\,}is a research scientist at Oak Ridge National Laboratory. He is the current Research Lead for the Center for AI Security (CAISER) at the lab investigating threats to and from AI. Amir received his Ph.D. from Cornell University, where his focus was on computer vision and machine learning. He is a member of IEEE.
\end{IEEEbiography}

\begin{IEEEbiography}{Cory Watson}{\,}is a Cyber Security Systems Engineer at Oak Ridge National Laboratory. He specializes in Linux system administration, network engineering, and software development with a general focus on supporting cybersecurity research. Cory is currently studying at Western Governor's University.
\end{IEEEbiography}

\begin{IEEEbiography}{Brian Weber}{\,}is a Cyber Research Engineer at Oak Ridge National Laboratory. Brian focuses on designing and building scalable software for conducting cybersecurity research, including a large scale COTS security tool evaluation framework, a scalable AI for cyber research platform and other projects. Brian received his M.S. from University of Maryland, Baltimore County in 2019.
\end{IEEEbiography}
\vspace*{8pt}
\end{document}